\documentclass[preprint2]{aastex}
\usepackage{natbib,amsmath}
\usepackage[bookmarks=true]{hyperref}

\begin{document}

\newcommand{\POET}{\emph{P.O.E.T.}}
\newcommand{\poet}{\emph{poet}}
\newcommand{\YREC}{\emph{YREC} } \newcommand{\ALGLIB}{\emph{ALGLIB} }
\newcommand{\briansrevisions}{} \newcommand{\kalosrevisions}{}

\title{POET: A Model for \textbf{P}lanetary \textbf{O}rbital
\textbf{E}volution due to \textbf{T}ides on Evolving Stars}
\author{Kaloyan Penev}
\affil{Department of Astrophysical Sciences, 4 Ivy Lane, Peyton Hall,
Princeton University, Princeton, NJ 08544, USA}
\author{Michael Zhang}
\affil{Department of Astrophysical Sciences, 4 Ivy Lane, Peyton Hall,
Princeton University, Princeton, NJ 08544, USA}
\author{Brian Jackson}
\affil{Carnegie DTM, 5241 Broad Branch Road, NW, Washington, DC 20015--1305,
USA}

\begin{abstract}
	We make publicly available an efficient, versatile, easy to use and
	extend tool for calculating the evolution of circular aligned planetary
	orbits due to the tidal dissipation in the host star. This is the first
	model to fully account for the evolution of the angular momentum of the
	stellar convective envelope by the tidal coupling, the transfer of
	angular momentum between the stellar convective and radiative zones, the
	effects of the stellar evolution on the tidal dissipation efficiency and
	stellar core and envelope spins, the loss of stellar convective zone
	angular momentum to a magnetically launched wind and frequency dependent
	tidal dissipation. This is only a first release and further development
	is under way to allow calculating the evolution of inclined and eccentric
	orbits, with the latter including the tidal dissipation in the planet and
	its feedback on planetary structure.  Considerable effort has been
	devoted to providing extensive documentation detailing both the usage and
	the complete implementation details, in order to make it as easy as
	possible for independent groups to use and/or extend the code for their
	purposes. POET represents a significant improvement over some previous
	models for planetary tidal evolution and so has many astrophysical
	applications. In this article, we describe and illustrate several key
	examples.
\end{abstract}

\section{Introduction}
The very first extrasolar (or exo-) planets discovered were planets with
orbital periods less than a few days, and the increasing number of such
planets over the last few decades shows they are not flukes -- close-in
exoplanets represent a robust outcome of planet formation and evolution.
Moreover, owing to observational biases, close-in exoplanets dominate
observational constraints on planetary composition, internal structure,
meteorology, etc., and so resolving the severe challenges they pose to
theories of formation is critical for extrapolating these constraints to all
planets.

Among other influences, tidal interactions between the planets and their host
stars shape the planetary population, and a long list of studies have
investigated the effects of tides in extrasolar systems, including
\citet{Sandquist_et_al_02, Sasselov_03, Levrard_et_al_09,
Jackson_Barnes_Greenberg_09, Debes_Jackson_10, Jackson_et_al_10,
Penev_et_al_12, Pepper_et_al_13, Collins_et_al_13, Zhang_Penev_13}. Close-in
exoplanets likely formed in more distant ($\gtrsim$ 1 AU), nearly circular
orbits, embedded in a protoplanetary disk well-aligned with the stellar
equator (\citealp{Lin_et_al_96}, although see \citealt{Batygin_Adams_13}),
and there are two standard scenarios for bringing them into close-in orbits:
(1) dynamical excitation of orbital eccentricity, followed by tidal evolution
and (2) gas disk migration.  Under scenario (1), gravitational excitations
scattered the planets into highly eccentric orbits, with pericenter distances
small enough for tidal interaction with the host star to bring the planets in
\citep{Rasio_Ford_96, Weidenschilling_Marzari_96, Fabrycky_Tremaine_07,
Nagasawa_et_al_08, Lithwick_Wu_13, Valeschi_Rasio_14}. Under scenario (2),
transfer of angular momentum from the planet's orbit to the protoplanetary
disk caused rapid inward migration \citep[e.g.][]{Chambers_09}.
\citet{Lin_et_al_96} suggested the migration would cease as the planet
entered a clearing in the gas disk near the host star, parking the planet in
a close-in orbit (at least temporarily).  Orbital decay from tides may then
continue, as long as the rotation of the host star lags behind the planet's
orbital motion.  \citet{Levrard_et_al_09} showed that most close-in gas
giants are unstable against tidal decay, and
\citet{Jackson_Barnes_Greenberg_09} showed their orbital distribution is
consistent with complete orbital decay and planetary disruption. More
recently, \citet{Teitler_Konigl_14} showed that the distribution of the
orbital and stellar rotation periods for short period planets discoverd by
the Kepler space telescope is consistent with tidal ingestion of planets,
which contributes to spinning up the host stars. A related topic,
\citet{Carlberg_et_al_09} showed that tidal decay during the post-main
sequence can even remove gas giants in more distant orbits resembling
Jupiter's.

Thus, tidal evolution spans from the pre- to post-main sequence, during which
time the host stars undergo significant evolution as well, and, as we show,
this stellar evolution can have dramatic consequences for the orbital
evolution of close-in exoplanets. However, all previous tidal studies
incorporated simplifying assumptions that exclude important effects. To
address this short-coming, we have developed and made publicly available the
\textbf{P}lanetary \textbf{O}rbital \textbf{E}volution due to \textbf{T}ides
(\POET) model. Using grids of stellar evolution tracks from the Yale Rotating
Stellar Evolution Code -- YREC -- \citep{Demarque_et_al_08}, \POET~is the
only presently available model that includes the coupling between stellar and
tidal evolution. 

Consequently, \POET~has a wide range of capabilities. For example,
\POET~tracks the stellar spin evolution, including angular momentum loss from
the stellar wind in a self-consistent way. \kalosrevisions{As we discuss in
Section \ref{sec: example calculations} (see Fig. \ref{fig: HAT-P-20}), this
can have profound consequences for the computed evolution. Unlike simplified
models which ignore stellar evolution \POET~is capable of computing the
evolution of orbits around pre-- and post--main sequence stars. An example of
this is again given in Section \ref{sec: example calculations} (see Fig.
\ref{fig: Kepler-91}).}

Since comparison of model predictions of tidal evolution to observations is
usually statistical in nature, we have made the code efficient, enabling
millions of model calculations with modest computational resources in a short
time.  Finally, ease of use has been a primary consideration. To that end, we
provide two interfaces to the tidal evolution calculator: a command line tool
and a python module, and both interfaces provide access to all model
parameters. We have also written extensive
\href{http://www.astro.princeton.edu/~kpenev/tidal\_orbital\_evolution/}
{documentation} \footnote{
\url{http://www.astro.princeton.edu/~kpenev/tidal_orbital_evolution/}}.

Of course, like any numerical model, \POET~has limitations. The most obvious
in the present version is the assumption that the planetary orbit is circular
and aligned with the stellar equator. Development is underway to relax these
assumptions for future versions of \POET. Of course we can never hope to
develop a completely universal tool, so we have devoted considerable effort
to document the code and make it as easy as possible for users to adapt it to
their applications.

In Sections \ref{sec: physics} and \ref{sec: numerics}, we briefly detail the
physical and numerical schemes employed by \POET, respectively. In Section
\ref{sec: stellar evolution}, we discuss the grid of \YREC stellar evolution
calculations. In Section \ref{sec: example calculations}, we present example
calculations, showing the importance of including the stellar rotation and
internal evolution. Finally, in Section \ref{sec: conclusions}, we discuss
limitations and planned improvements to the code. Appendix \ref{app:
installation} explains how to install \POET~Appendices \ref{app: stopping
conditions} and \ref{app: stellar evolution} provide a more detailed
description of the numerical scheme.

\section{Physical Model}
\label{sec: physics}

We define our notation up front (see text for details):

\begin{description}
\item[$M_*$] mass of the star
\item[$R_*$] radius of the star
\item[$m_p$] mass of the planet
\item[$r_p$] radius of the planet
\item[$Q_*$] modified tidal quality factor of the star
	\citep[cf.][]{Ogilvie_Lin_07}
\item[$a$] semimajor axis of the orbit
\item[$\omega_\mathrm{surf}$] angular velocity of the stellar surface
	($\omega_\mathrm{conv}$ for low mass stars and the solid body rotation
	for high mass stars)
\item[$\omega_\mathrm{conv/rad}$] angular velocity of the stellar
	convective/radiative zone
\item[$\omega_\mathrm{orb}$] orbital angular velocity
\item[$I_\mathrm{conv/rad/*}$] moment of inertia of the stellar
	convective/radiative zone or the entire star.
\item[$L_\mathrm{conv/rad}$] angular momentum of the stellar
	convective/radiative zone
\item[$L_*$] angular momentum of an entire high mass star
\item[$K$] parameter giving the strength of the magnetic wind of the star 
\item[$\omega_\mathrm{sat}$] stellar surface angular velocity at which the
	magnetic wind saturates
\item[$\tau_c$] stellar core--envelope coupling timescale
\item[$M_\mathrm{rad}$] mass of the radiative core for low mass stars
\item[$R_\mathrm{rad}$] radius of the radiative--convective boundary in low
	mass stars
\item[$M_*^\mathrm{crit}$] stellar mass below which stars are split into a
	radiative core and a convective envelope and above which they are treated
	as solid bodies, ignoring the tiny surface convective zone.
\end{description}

Our model computes the secular evolution of a planet--star system as its
semi--major axis evolves due to the dissipation of the tides raised by the
planet on the star. In fact the dissipation of the tides raised on the planet
by the star may also be important. However, since the angular momentum of the
planetary spin is very small compared to the angular momenta of the orbit or
the star, the planet's rotation is synchronized quickly with the orbit. For
an eccentric orbit, this would still lead to time dependent tides on the
planet, however, our model currently assumes circular orbits. As a result,
once the planet's rotation is synchronized to the orbit the planetary tides
are stationary and not subject to dissipation.

Under these assumptions, the evolution of the semi--major axis is given by
\citep{Goldreich_63, Kaula_68, Jackson_et_al_08a}:
\begin{equation}
	\dot{a}=\mathrm{sign}(\omega_\mathrm{surf}-\omega_\mathrm{orb})
	\frac{9}{2}\sqrt{\frac{G}{aM_*}}\left( \frac{R_*}{a}
	\right)^5\frac{m_p}{Q_*}\quad \label{eq: adot}\\
\end{equation}

Where $\mathrm{sign}(\omega_\mathrm{surf}-\omega_\mathrm{orb})$ takes the
value 1 when the stellar surface is spinning faster than the planet
and -1 when it is spinning slower. 

This expression neglects the planet's mass relative to the star's mass, a
perfectly reasonable assumption, given the uncertainties, and indeed the lack
of a good physical model for $Q_*$.

Angular momentum conservation requires that any angular momentum gained or
lost by the orbit is taken from or added to the star, changing its spin.
Simple arithmetic shows that the rate at which angular momentum is deposited
into the star due to the orbit evolving is given by:
\begin{equation}
	\dot{L}_\mathrm{tide}\equiv-\frac{1}{2}m_p
	M_*\sqrt{\frac{G}{a(M_*+m_p)}}\dot{a}
	\label{eq: Ldot_tide}
\end{equation}

Here we do not neglect the mass of the planet relative to the stellar mass,
in order to make angular momentum conservation exact. Among other things,
this allows users to judge if the precision requirements they have specified
is sufficient for the evolution they are calculating.

Our model makes a distinction between low mass stars, which have appreciable
surface convective zones, and high mass stars, which do not.

For low  mass stars ($M_*<M_*^\mathrm{crit}$), the tidal dissipation is
assumed to occur in the surface convective zone, and the angular momentum is
deposited only in that zone, while for high mass stars no such splitting is
made.

Furthermore, measurements of stellar spins in open clusters reveal that low
mass stars lose most of their initial angular momentum over their lifetime.
This is thought to be due to stellar winds which are coupled to the star via
its magnetic field. This effect has been extensively studied
\citep[c.f.][among many others]{Schatzman_62, Skumanich_72, Kawaler_88,
Irwin_Bouvier_09, Gallet_Bouvier_13}, but remains poorly understood. A
combination of theory and observation motivates our formulation of these
effects \citep{Stauffer_Hartmann_87, Kawaler_88, Barnes_Sofia_96}:
\begin{equation}
	\dot{L}_\mathrm{wind}\equiv -K\omega_\mathrm{surf}
	\min(\omega_\mathrm{surf}, \omega_\mathrm{sat})^2 \left(
	\frac{R_*}{R_\odot}\right)^{1/2} \left( \frac{M_*}{M_\odot}
	\right)^{-1/2} \label{eq: Ldot_wind}
\end{equation}

Finally, \citet{Irwin_et_al_07, Irwin_Bouvier_09, Denissenkov_10}, argue that
in order to explain the observed rotation rates in open clusters of different
ages, it is necessary to allow the cores and envelopes of low mass stars to
spin at different rates but that they are coupled on a timescale of a few Myr.
Our model uses a formulation by \citet{MacGregor_91} and \citet{Allain_98}
according to which the rate of angular momentum transfer from the core to the
envelope is given by:
\begin{equation}
	\dot{L}_\mathrm{coup}\equiv\frac{\Delta L}{\tau_c} - \frac{2}{3}
	R_\mathrm{rad}^2 \omega_\mathrm{conv} \dot{M}_\mathrm{rad}
	\label{eq: Ldot_coup}
\end{equation}
where
\begin{equation}
	\Delta L\equiv\frac{I_\mathrm{conv}L_\mathrm{rad}-
	I_\mathrm{rad}L_\mathrm{conv}}{I_\mathrm{conv}+I_\mathrm{rad}} \label{eq:
	DeltaL}
\end{equation}

For high mass stars ($M_*>M_*^\mathrm{crit}$), observations suggest that the
angular momentum loss is not important \citep[c.f.][]{Kraft_67,
Zorec_Royer_12}. Nonetheless, in the interest of generality, we still use Eq.
\ref{eq: Ldot_wind} to describe angular momentum loss for high mass stars,
albeit with different value of $K$, which can be set to zero if no angular
momentum loss should occur. 

In our model, high mass stars are not split into zones, but rather solid body
rotation is assumed for the star. As a result the rotational evolution
simplifies to:

\begin{equation}
	\dot{L}_*=\dot{L}_\mathrm{wind} + \dot{L}_\mathrm{tide} \label{eq: L*dot}
\end{equation}

Finally, we follow the evolution of the stellar quantities ($R_*$,
$I_{conv}$, $I_{rad}$, $R_{rad}$ and $M_{rad}$) as the stellar structure
evolves. Further, $Q_*$ can be an arbitrary function of the tidal frequency
(the difference between the orbital and stellar spin frequencies). Details on
how users can set the frequency dependence of $Q_*$ are given in Appendix
\ref{app: installation}.
As described in the next section, these basic equations are re-formulated and
combined in some cases to facilitate and stabilize the numerical scheme.

\section{Numerical Scheme}
\label{sec: numerics}

\subsection{Stopping Conditions}
Because Equations \ref{eq: adot} and \ref{eq: Ldot_wind} have discontinuities
(when $\omega_\mathrm{surf}=\omega_\mathrm{orb}$ for Eq. \ref{eq: adot} and
when $\omega_\mathrm{surf}=\omega_\mathrm{sat}$ for Eq. \ref{eq: Ldot_wind}),
it is beneficial to detect when the evolution goes through these
discontinuities and ensure that it does not simply jump over such points, but
lands exactly on them (to some precision of course). Such special treatment
allows the evolution to be calculated both more accurately and more
efficiently.

To see this, consider the discontinuity in Eq. \ref{eq: adot}. Because the
sign of the tidal evolution changes when the spin of the star goes through
synchroneity with the orbit, it is possible to lock the system in a state
where $\omega_\mathrm{surf}$ is held equal to $\omega_\mathrm{orb}$. If we
simply let the ordinary differential equation (ODE) solver handle this for
us, the best possible outcome would be to take tiny steps, oscillating
between super-- and sub--synchronous rotation. However, if we go through the
effort of detecting this and stopping the evolution precisely at the point
where synchronous rotation is achieved, we can switch to a different system
of differential equations that assumes a spin--orbit lock and uses it to
eliminate one of the evolution variables.  This avoids the oscillatory
behavior, and large time steps can safely be taken.

The spin--orbit lock may not persist indefinitely. The orbit continues to
evolve since the system is losing angular momentum due to the stellar wind.
Consequently, there may come a point when the tidal dissipation in the star
cannot drain sufficient amount of angular momentum from the orbit to
compensate for the wind loss and the extra spin up required of the star in
order to match the shorter orbital period, at which point the evolution has
to revert back to the non--locked equations.

Next, consider the discontinuity in Eq. \ref{eq: Ldot_wind}. Because in this
case the rate of evolution of the angular momentum of the convective region
(or total angular momentum for the case of a high mass star) is not
discontinuous, but its derivative is, the ODE solver can blindly jump over
the $\omega_\mathrm{surf}=\omega_\mathrm{sat}$ point resulting in the change
from saturated to non--saturated wind (or vice--versa) happening later than
it should. If on the other hand, we detect this and force the solver to land
precisely on the critical point, the calculated evolution will be more
precise.

In addition to the above discontinuities, we have several others, which are
due to the fact that we may want to include parts of the evolution of the
system before and after the planet is present.

In the present version of the code, we are able to start the evolution when
the protoplanetary disk is still present, assuming that the stellar surface
rotation is locked to the rotation rate of the inner edge of the disk
\citep{Lin_et_al_96}. Then at some specified age, the disk is removed
(releasing the surface rotation rate of the star from the lock) and replaced
with a planet in a circular orbit. We follow the evolution until either the
star leaves the main--sequence or the orbit shrinks so much that the planet
is tidally disrupted or engulfed by the star, a condition we refer to as
``planet death''.

The semimajor axis at which the planet is assumed to be
tidally disrupted is calculated according to \citet{Roche_50, Roche_51}:
\begin{equation}
	a_{roche}=2.44r_p\left(\frac{M_*}{m_p}\right)^\frac{1}{3}
\end{equation}
and the planet is assumed to be engulfed by the star when the semimajor axis
is equal to the stellar radius.

If the planet dies before the star leaves the main--sequence, the angular
momentum of the planetary orbit at the moment of death is added to the
stellar convective zone for low mass stars or to the angular momentum of the
entire star for high mass stars. We then follow the subsequent rotational
evolution of the star until the end of its main sequence phase.

In addition, we allow for user-defined stopping conditions (some function of
the orbital parameters and age which are either interesting to the user or
indicate that the evolution equations must be modified). Each Stopping
Condition should be a quantity that varies smoothly with the evolution
(continuous and continuously differentiable up to at least third order) and
is zero exactly when the evolution should be stopped. See Appendix \ref{app:
stopping conditions} for a detailed description of how stopping conditions
are handled. Pre--defined stopping conditions handle all the discontinuities
in the evolution equations, and users can define additional stopping
conditions as functions of the orbital and system parameters, without needing
to understand any of the implementation details. Appendix \ref{app:
installation} gives details on how that can be achieved.

\subsection{Evolution Modes}
As discussed above, discontinuities in the evolution require switching
between different systems of differential equations when some
\verb#StoppingCondition# is encountered. The system of differential equations
to use at any given time is determined by an evolution mode and the wind
saturation state.

\subsubsection{Wind Saturation States}
\label{sec: wind saturation modes}
The wind saturation state only affects how $\dot{L}_\mathrm{wind}$ is
calculated.\\

\emph{NOT\_SATURATED}
\begin{equation}
	\dot{L}_\mathrm{wind}=-K\omega_\mathrm{surf}^3
		\left(
			\frac{R_*}{R_\odot}\right)^{1/2} \left( \frac{M_*}{M_\odot}
		\right)^{-1/2}
\end{equation}\\

\emph{UNKNOWN}
\begin{equation}
	\dot{L}_\mathrm{wind}=-K\omega_\mathrm{surf}
			\min(\omega_\mathrm{surf}, \omega_\mathrm{sat})^2 
		\left(
			\frac{R_*}{R_\odot}\right)^{1/2} \left( \frac{M_*}{M_\odot}
		\right)^{-1/2}
\end{equation}\\

\emph{SATURATED}
\begin{equation}
	\dot{L}_\mathrm{wind}=-K\omega_\mathrm{surf}\omega_\mathrm{sat}^2 
		\left(
			\frac{R_*}{R_\odot}\right)^{1/2} \left( \frac{M_*}{M_\odot}
		\right)^{-1/2}
\end{equation}

The \emph{UNKNOWN} state never actually occurs during evolution, but is useful
when defining stopping conditions for example.

\subsubsection{Stellar Rotation and Orbital Evolution Modes}
Depending on the stellar rotation evolution mode, the rotational and orbital
evolution are calculated using different sets of variables and equations
governing their evolution.\\

\emph{LOCKED\_TO\_DISK}

This is the evolution mode for a system for which the protoplanetary disk is
still present. In this case, the spin of the surface convective zone is held
at some prescribed constant value $\omega_\mathrm{disk}$, representing the
orbital frequency of the inner edge of the protoplanetary disk, to which the
stellar surface rotation is locked.

In this case, the only evolution variable is $L_\mathrm{rad}$, and the
equation governing its evolution is:
\begin{eqnarray}
	\dot{L}_\mathrm{rad}&=&-\dot{L}_\mathrm{coup}
\end{eqnarray}

Where $L_\mathrm{conv}$ in Eq. \ref{eq: Ldot_coup} is replaced by
$I_\mathrm{conv}\omega_\mathrm{disk}$.

The reason for including this extra stage is that it makes it possible to
start the evolution at the age when the radiative core first begins to form,
thus not requiring an initial value to be supplied for its angular momentum
(or spin frequency). Instead it acquires angular momentum through
core--envelope coupling.

The evolution will switch out of this mode at a prescribed disk--dissipation
age. The subsequent evolution mode is FAST\_PLANET, LOCKED\_TO\_PLANET or
SLOW\_PLANET depending on the initial semimajor axis at which the planet
appears.

This evolution mode only makes sense for low mass stars, since for high mass
stars there are no variables left to evolve.\\

\emph{FAST\_PLANET}

This is the evolution mode for a system in which the orbital period is
shorter than the spin period of the stellar surface. In this case the
evolution variables are: $a^{6.5}$, $L_\mathrm{conv}$ and $L_\mathrm{rad}$
for low mass stars and $a^{6.5}$ and $L_*$ for high mass stars. The equations
for their evolution are:
\begin{eqnarray}
	\frac{da^{6.5}}{dt}&=&
	-\frac{117}{4}\sqrt{\frac{G}{M_*}}R_*^5\frac{m_p}{Q_*}\\
	\dot{L}_\mathrm{conv}&=&\dot{L}_\mathrm{coup} + \dot{L}_\mathrm{wind} +
	\dot{L}_\mathrm{tide}\label{eq: fast Lconv dot}\\
	\dot{L}_\mathrm{rad}&=&-\dot{L}_\mathrm{coup}\label{eq: fast Lrad dot}\\
	\dot{L}_*&=& \dot{L}_\mathrm{wind} + \dot{L}_\mathrm{tide} \label{eq:
	fast L* dot}
\end{eqnarray}
where Equations \ref{eq: fast Lconv dot} and \ref{eq: fast
Lrad dot} are used for low mass stars and eq. \ref{eq: fast L* dot} for high
mass stars.

The reason for using $a^{6.5}$ instead of $a$ as the evolution variable is
evident from the first equation above. The rate at which $a^{6.5}$ evolves is
independent of $a$. In fact, for a constant $Q_*$ it only changes due to
$R_*$ evolving. This allows the ODE solver to take much larger steps when the
orbit has shrunk than would otherwise be possible. 

This evolution mode can end in one of two ways:
\begin{enumerate}
	\item The planet dies, and the subsequent evolution mode is NO\_PLANET.
	\item The spin period of the stellar surface matches the orbital period,
		in which case the subsequent evolution mode is either
		LOCKED\_TO\_PLANET or SLOW\_PLANET, depending on whether the transfer
		of angular momentum due to tides is sufficient to keep the lock.
\end{enumerate}

\emph{LOCKED\_TO\_PLANET}

This is the evolution mode for a system in which the surface rotation of the
star is held locked to the orbit by the dissipation of the stellar tides.

For low mass stars, the evolution variables are $a$ and $L_\mathrm{rad}$ and
the equations:
\begin{eqnarray}
	\dot{a}&=&2\frac{T - \dot{I}_\mathrm{conv}/a} {M_*m_p/(M_*+m_p) -
	3I_\mathrm{conv}/a^2}\\
	T&\equiv&\sqrt{\frac{a}{(M_*+m_p)G}}\left[\dot{L}_\mathrm{coup} +
	\dot{L}_\mathrm{wind}\right]\\
	\dot{L}_\mathrm{rad}&=&-\dot{L}_\mathrm{coup}
\end{eqnarray}

For high mass stars, the only variable is $a$ and it is evolved according to:
\begin{eqnarray}
	\dot{a}&=&2\frac{T - \dot{I}_*/a} {M_*m_p/(M_*+m_p) - 3I_*/a^2}\\
	T&\equiv&\sqrt{\frac{a}{(M_*+m_p)G}}\dot{L}_\mathrm{wind}\\
\end{eqnarray}

This evolution mode can end either by the planet dying or by the rate of
transfer of angular momentum from the orbit to the star falling below what is
required to keep the lock. In the first case, the subsequent evolution mode
is NO\_PLANET and in the other case it is either FAST\_PLANET or
SLOW\_PLANET.\\

\emph{SLOW\_PLANET}

This is the evolution mode for a system in which the orbital period is longer
than the spin period of the stellar surface convective zone. In this case the
evolution variables are the same as for the FAST\_PLANET case: $a^{6.5}$,
$L_\mathrm{conv}$, $L_\mathrm{rad}$ for low mass stars and $a^{6.5}$, $L_*$
for high mass stars, and the equations for their evolution are identical,
except for a sign change in the equation for the evolution of the semimajor
axis.

This evolution mode can end only if The spin period of the stellar surface
convective zone matches the orbital period. In which case the subsequent
evolution mode is either LOCKED\_TO\_PLANET or FAST\_PLANET, depending on
whether the transfer of angular momentum due to tides is sufficient to keep
the lock.\\

\emph{NO\_PLANET}

This is the evolution mode for a star without a planet in orbit and no
protoplanetary disk. Usually this state is reached after the planet dies.
The evolution variables are $L_\mathrm{conv}$ and $L_\mathrm{rad}$ for low
mass stars and $L_*$ for high mass stars. Their evolution is given by:
\begin{eqnarray}
	\dot{L}_\mathrm{conv}&=&\dot{L}_\mathrm{coup} + \dot{L}_\mathrm{wind}\\
	\dot{L}_\mathrm{rad}&=&-\dot{L}_\mathrm{coup}\\
	\dot{L}_*&=&\dot{L}_\mathrm{wind}
\end{eqnarray}

This evolution mode persists until the end of the star's lifetime.

\begin{figure*}[t!]
	\begin{center}
		\includegraphics[width=\textwidth]{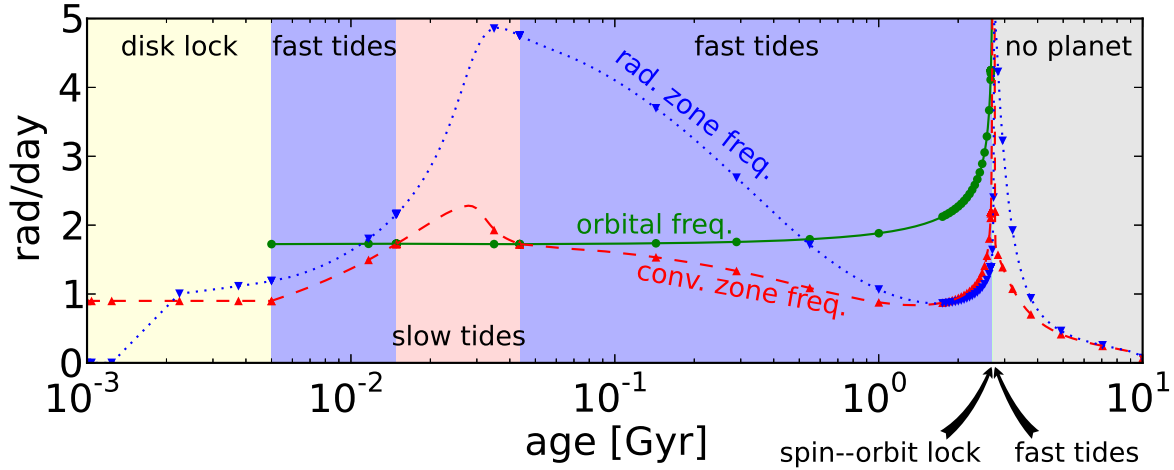}
		\caption{\footnotesize The orbital evolution of a 10 Jupiter mass
		planet around a solar mass star. The lines show the evolution with an
		imposed maximum time step of 0.1 Myr and the symbols show the time
		steps taken by our code when no limit is imposed. The line styles and
		symbols show the evolution of the orbital angular velocity (solid
		green line and green circles), the angular velocity of the convective
		envelope of the star (dashed red line and red upward triangles) and
		the angular velocity of the radiative core of the star (dotted blue
		line and blue downward triangles). The background denotes the various
		evolution modes detected by \POET~From left to right: the convective
		zone is locked to a disk with a prescribed angular velocity; the
		planet forms in an orbit faster than the surface spin of the star; the
		star spins up due to shrinking, exceeding the orbital angular
		velocity; the stellar wind removes sufficient angular momentum from
		the star so that its surface rotation drops below the orbital
		frequency again; the transfer of angular momentum from the orbit to
		the star synchronizes and locks the surface rotation of the star to
		the orbit; the lock is broken and the star spins slower than the
		planet, and finally the planet spirals into the star and deposits its
		angular momentum resulting in very fast stellar spin which decays due
		to the stellar wind.} \label{fig: example orbit}
	\end{center}
\end{figure*}

\section{Stellar Evolution}
\label{sec: stellar evolution}

Because the goal of \POET~is to calculate evolution as efficiently as possible,
stellar evolution is handled by interpolating among a pre--computed grid of
tracks based on the \YREC stellar evolution model \citep{Demarque_et_al_08}.
In addition, there is a mechanism for users to supply a custom stellar
evolution track.

The built-in \YREC~tracks are for solar metallicity stars with masses
0.5, 0.6, 0.7, 0.8, 0.9, 1, 1.05, 1.1, 1.15 and 1.2 solar masses.  There are
two main limitations to this grid: no track extends beyond an age of 10 Gyr
or past the point that the star turns off the main sequence. Also, tracks are
available only for solar metallicity stars.  For this grid, the value for a
particular stellar quantity at a given age and stellar mass is
calculated by first interpolating each of the grid tracks to a scaled value
of the desired age and then interpolating among the values to the desired
stellar mass. For a complete description of the interpolation algorithm, see
Appendix \ref{app: stellar evolution}.

We employ cubic splines from the
\href{http://www.alglib.net/interpolation/leastsquares.php#splinefit}
{\ALGLIB} library\footnote{
\url{http://www.alglib.net/interpolation/leastsquares.php\#splinefit}} to
smooth the evolutionary tracks and tame numerical artifacts that would
otherwise result from our use of second derivatives. We then interpolate
between tabulated time steps in the stellar evolution tracks to the desired
times. While this approach results in robust and fast interpolation, it takes
quite a long time to actually derive the smoothing splines. In order to avoid
this overhead every time \POET~is run, the interpolation can be derived once
for a grid and ``serialized'' (saved) to a file using the
\verb#boost_serialization# library (see Appendix \ref{app: installation}) for
future re--use.

It is possible for users to provide an individual track to use as the stellar
evolution model if their application requires it. The track is assumed to be
applicable for all stars in a run, and only interpolation in age is used.
Since, similarly to the tracks included with \POET, numerical artifacts can
result in unusable interpolations, care must be taken when deriving
interpolations for user supplied tracks. The best policy is to output and
examine the interpolated quantities as well as their first and second
derivatives by eye, and adjust the smoothing parameters for the interpolation
if necessary.

User supplied tracks can be ``serialized'' for future re--use just like the
built--in grids.

\section{Example Calculations}
\label{sec: example calculations}

Including the evolution of the stellar interior and rotation in a
self-consistent way results in much richer behavior, depending on the
parameters of the system being evolved. As an illustration, in Fig. \ref{fig:
example orbit} we show the evolution of a 10 Jupiter mass planet around a
solar mass star. The values of $Q_*$ and the initial semimajor axis were
chosen such that the resulting evolution goes through all evolution modes.
The stellar wind and core--envelope coupling were left at their default
values.  Both the lines and points were calculated using \POET~The only
difference is that for the lines the maximum time steps was limited to 0.1
Myr, while the symbols show the optimal time steps chosen by the code with no
limit imposed, thus demonstrating the large time steps \POET~is able to take
without sacrificing accuracy in the computed evolution. As a reference,
computing the evolution with automatic step size control required about one
second (on a run--of--the--mill desktop), most of which was taken up by
reading in the ``serialized'' stellar evolution), while the finely sampled
calculation took close to 20 seconds, clearly dominated by actually computing
the evolution.

Following the evolution of the stellar properties along with the evolution of
the orbit allows studies that are otherwise impossible.
\citet{Penev_et_al_12} modelled detection probabilities for tidally evolving
extrasolar planets around evolving stars during the whole stellar main
sequence. This calculation incorporated detection biases and provided limits
on the tidal dissipation parameter $Q_*$ for the host stars from the observed
distribution of exoplanet orbital periods.

\citet{Zhang_Penev_13} suggested using the fact that stars spin up to
extremely short periods if they accrete a hot Jupiter to select candidate
stars for which this may have occurred. The ability to follow the evolution
of the stellar spin was clearly crucial in estimating the prevalence of such
fast rotators.

\begin{figure}
	\begin{center}
		\includegraphics[width=0.45\textwidth]{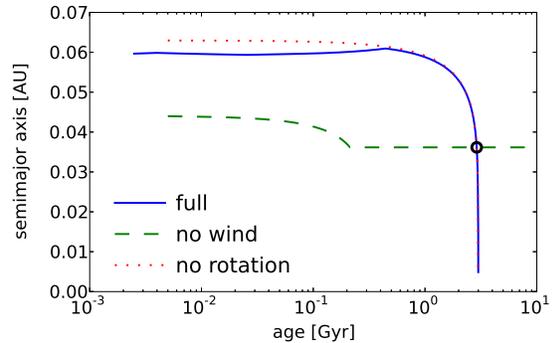}
		\caption{The evolution of HAT-P-20b's orbit for $Q_*=10^5$ including
		all the details \POET~is capable of following (solid blue curve),
		ignoring the evolution of the star, the magnetic wind and
		core--envelope decoupling (dashed green curve) and fully ignoring any
		evolution and assuming no rotation for the parent star (dotted red
		curve). The black circle where all three lines intersect denotes the
		observed orbit at the assumed present age of the system.} 
		\label{fig: HAT-P-20}
	\end{center}
\end{figure}

Including the stellar evolution is particularly important for systems like
HAT-P-2 \citep{Bakos_et_al_07a}, HAT-P-20 and HAT-P-21
\citep{Bakos_et_al_11}, and WASP-10 \citep{Christian_et_al_09}, for which the
planets' orbital angular momenta are comparable to the host stars' spin
angular momentum. For these cases, tidal interactions can temporarily
synchronize the stellar spins before the planets finally plunge into their
host stars. In Fig. \ref{fig: HAT-P-20} we show the evolution of the orbit of
HAT-P-20b for $Q_*=10^5$ calculated using three sets of assumptions:

\begin{figure}
	\begin{center}
		\includegraphics[width=0.45\textwidth]{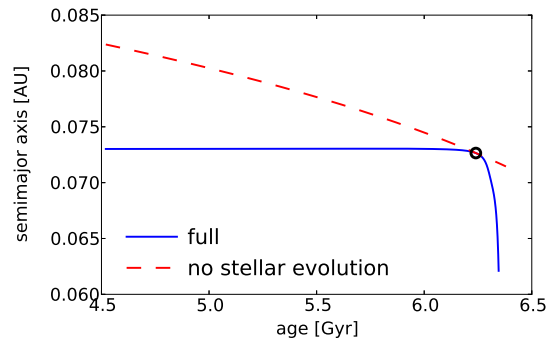}
		\caption{The evolution of KOI-2133.01's orbit for $Q_*=10^8$ including
		all the details \POET~is capable of following (solid blue curve), and
		ignoring the evolution of the star (dashed red curve). The black
		circle where the two lines touch denotes the observed orbit at the
		assumed present age of the system.}
		\label{fig: Kepler-91}
	\end{center}
\end{figure}

\begin{table*}
	\begin{center}
		\begin{tabular}{r|c|c|c|c|c}
			& \multicolumn{3}{c}{HAT-P-20b}	& \multicolumn{2}{c}{KOI-2133}\\
			& full & no wind & no rotation & full & no stellar evolution \\
			\hline
			$K$ [$M_\odot R_\odot^2 \mathrm{day}^2 \mathrm{rad}^{-2}
			\mathrm{Gyr}^{-1}$]		& 0.17	& 0		& $\infty$ & 0 & 0\\
			$\omega_{sat}$ [rad/day]		& 2.45	& 2.45	& 2.45		& N/A
			& N/A\\
			$\tau_c$ [Myr]					& 28	& 0		& 0			& N/A
			& N/A\\
			$Q_*$							& $10^5$& $10^5$& $10^5$ &
			$10^8$&$10^8$\\
			$M_*$ [$M_\odot$]				& 0.756	& 0.756	& 0.756 & 1.31	&
			1.31\\
			$m_p$ [$M_\mathrm{jup}$]		& 7.3	& 7.3	& 7.3 & 0.88	&
			0.88\\
			$r_p$ [$R_\mathrm{jup}$]		& 0.876	& 0.876	& 0.876 & 1.384 &
			1.384\\
			initial stellar spin period [days]	& 7	&$\infty$& $\infty$ & 6.0
			& $\infty$\\
			planet appearance age [Myr]		& 5		& 5		& 5 & 4520	&
			4520 \\
			initial orbital period [days]	&5.96383&3.85693& 6.604712 &
			6.29362 & 7.541190\\
		\end{tabular}
	\end{center}
	\caption{The values for the various model parameters used to calculate
	the evolutions of HAT-P-20b's and KOI-2133's orbits}
	\label{tbl: initial conditions}
\end{table*}

\begin{itemize}
	\item the full evolution of the stellar structure and rotation, including
		angular momentum loss to the stellar wind, asynchronous rotation
		between the stellar radiative core and convective envelope, and
		starting the star with reasonable initial rotation (solid blue
		curve).
	\item assuming solid body rotation for the star, ignoring the loss of
		angular momentum to stellar wind, and starting the star without any
		rotation at 5Myr (dashed green curve).
	\item ignoring both the rotation and the evolution of the star, and
		simply calculating the evolution of the semimajor axis (dotted red
		curve).
\end{itemize}
In each case, the initial conditions were tuned in order to make the orbital
period attain its presently observed value at a system age of 2.9 Gyr (see
table \ref{tbl: initial conditions}). The age was chosen because at that age
our stellar evolution reproduces the nominal radius of HAT-P-20, and it is
consistent (within the uncertainties) with the value quoted in
\citet{Bakos_et_al_11}.

The simplest possible assumptions (assuming nothing about the star changes)
do very well at reproducing the future HAT-P-20 orbit. This is because, even
though the orbit has much more angular momentum than necessary in order to
synchronize the star, after synchronization, the star spins so fast, that it
loses angular momentum to its wind at a very high rate. As a result, the
tidal dissipation we assumed is just barely large enough to hold the
spin--orbit lock, and the orbital evolution proceeds at almost the same rate
as under the assumption of a non-rotating star. The early evolution is quite
different, due to the fact that for the full evolution, for ages between 25
and 440 Myr, the stellar spin period is shorter than the orbital period (due
to the star shrinking onto the main sequence). This results in the planet
actually being pushed away from the star during this period.

This calculation illustrates an important result: even after tidal
interactions synchronize the spin of a host star, the planet may still die.
\citet{Levrard_et_al_09} pointed out that planetary systems with sufficiently
large total angular momenta (orbital + spin angular momenta) are formally
stable against tidal decay. However, as suggested by \citet{Levrard_et_al_09,
Barker_Ogilvie_09}, the continual loss of angular momentum through the
stellar wind, included in this study, may doom such planetary systems anyway.
In fact, since the rate of angular momentum loss increases with stellar spin,
spin synchronization usually results in a tidal decay rate very similar to
that computed by assuming the star is not rotating.

Another class of systems which absolutely require the simultaneous
calculation of the stellar evolution along with the orbit are planets around
evolved stars, like HD 102956b \citep{Johnson_et_al_10}, HIP 13044b
\citep{Klement_et_al_11} and KOI-2133.01 \citep{Lillo-Box_et_al_13}. For
these systems, the timescale on which the parent star evolves is comparable
to, or even shorter than the timescale for orbital evolution. 

Figure \ref{fig: Kepler-91} demonstrates the difference between including and
not the stellar evolution when calculating the orbital evolution of
KOI-2133.01. Since the built-in YREC tracks do not go past the end of the
main sequence lifetime of any star and do not extend to masses as high as
1.31 (KOI-2133's mass), we used MESA \citep{MESA_1, MESA_2} to generate a
short post--main--sequence track suitable for this system, which was passed
to \POET~ as a custom stellar evolution. We found that the stellar properties
quoted in \citet{Lillo-Box_et_al_13} match our track best for an age of 6.24
Gyr (consistent with the system age quoted in the paper to within the
uncertainty), so we adopted a present system age of 6.24 Gyr. As before, we
tuned the initial conditions (see table \ref{tbl: initial conditions}) to
match the presently observed state of the system. As expected, the simplified
calculation over--predicts the rate of orbital evolution at early times,
because it overestimates the radius of the star, and under--predicts the
evolution at later times, since it underestimates the stellar radius.

\section{Conclusions}
\label{sec: conclusions}

We have presented a code (\POET) capable of calculating the evolution of
extrasolar planet orbits taking into account the rotation and the evolution
of the structure of the parent star, including the transfer of angular
momentum from the orbit to the star and the loss of angular momentum by the
star to a magnetically launched wind. 

Because it properly handles all these effects, unlike previous models,
\POET~is capable of following the evolution of planetary orbits from before
the star lands on the main sequence to well after it leaves it. Also it
properly takes into account the possibility that planets may spin--up their
star to synchronous rotation with the orbit, but that this does not mean that
further evolution stops.

As any model, \POET~has limitations. First, at present \POET~is limited to
calculating the evolution of only circular orbits perfectly aligned with the
stellar equator. This precludes investigations into some of the formation
scenarios for hot Jupiters, which predict that exoplanet orbits start with
significant eccentricities and/or are significantly inclined with respect to
the stellar equator. We plan to address this in two steps: i) implement
evolving inclined orbits following the formalism of \citet{Lai_2012}, ii)
introduce eccentricity, by extending that formalism. The first step is a
relatively straight forward extension of the current implementation, which
simply requires adding more parameters (the seven dissipation efficiencies,
or equivalently tidal quality factors, introduced in \citet{Lai_2012}. The
second step is more involved, since for eccentric orbits the planet
experiences time variable tidal forces, no matter its spin. This means that a
complete calculation must include the dissipation in the planet as well as
the star. Further, since unlike for stars, the rate at which energy is
deposited in a planet by the tidal dissipation may be important, or even
completely dominate the energy budget of the planet, its effects on the
planetary structure must be included (cf. \citealp{Miller_et_al_09}).

Second, the grid of stellar evolution models presently included is limited.
On the one hand, no tracks extend past the end of the main sequence, and in
fact for stars with mass lower than 1 $M_\odot$ tracks are terminated at 10
Gyr.  Further, only tracks for solar metallicity stars are available. Since
we provide a mechanism for users to supply their own stellar evolution
tracks, these limitations can be easily overcome, as we demonstrated by
calculating the evolution of KOI-2133.01's orbit (see Sec. \ref{sec: example
calculations}). Nevertheless, We are currently in the process of generating
more extended grids of stellar evolution models using the MESA suite, which
will also be made publicly available.

\bibliography{bibliography}
\bibliographystyle{apj}

\appendix

\section{Installation and Usage}
\label{app: installation}

The latest official release of \POET~can be downloaded from:\\

\begin{center}
\url{http://www.astro.princeton.edu/~kpenev/tidal_orbital_evolution/poet.tgz}.
\end{center}
or
\begin{center}
\url{https://www.assembla.com/spaces/tidal-orbital-evolution}
\end{center}

In addition, the python module is installable from
\href{https://pypi.python.org/pypi}{PyPI -- the Python Package Index}
\footnote{\url{https://pypi.python.org/pypi}} (package name \verb#POET#). The
\verb#PyPI# website provides full instructions on how to install packages
hosted there.

Compilation requires the development version (header files as well as shared
library files) of the following libraries: \verb#gsl#\footnote{\url{http://www.gnu.org/software/gsl/}}, \verb#boost_serialization#\footnote{\url{http://www.boost.org/}},
and \verb#argtable2#\footnote{\url{http://argtable.sourceforge.net/}}. Those
are available as packages on most Linux and Mac (through \verb#macports#)
distributions.  All other libraries needed by the code are already directly
included with the source.

After those dependencies are met, compiling and installing the code is
simple. After changing to the \verb#src# sub--directory, run:
\begin{verbatim}
	sudo make install BINDIR='<desired location of executables>' DATADIR='<desired location of data files>'
\end{verbatim}
This will compile an executable called \verb#poet# and copy it to the
destination you specified with the \verb#BINDIR# option, and copy \POET~
related data (e.g. stellar evolution tracks) to the destination specified
with \verb#DATADIR#. Finally, it will compile and install in the standard
location for your system a python module named \verb#poet#.

For a full documentation of the available \verb#make# targets run 
\verb#make help# or see:\\

\url{http://www.astro.princeton.edu/~kpenev/tidal_orbital_evolution/compilation.html}.
\\

Successful compilation produces a single executable named \poet~and places it
in the location specified by the \verb#BINDIR# argument to
\verb#make install#. In addition, \POET~related data is copied to the
location specified by the \verb#DATADIR# argument. Finally, a python module
named \verb#poet# is compiled and installed in the standard location for
your system. Commonly, you would want to make sure that \verb#BINDIR# is in
your search path.

All the parameters that enter into the evolution equations (see Section
\ref{sec: physics}) are changeable through command line options of
\poet. Rather than listing here the close to 50 command line options,
which may change in the future, we outline the general scheme used and for
details refer the reader to:\\

\url{http://www.astro.princeton.edu/~kpenev/tidal_orbital_evolution/usage.html},
\\

which always documents the latest release. The same information
can also be obtained by invoking \verb#poet -h#.

In order to fully define the problem, many parameters need to have their
values specified. In addition, many applications will require calculating a
large number of orbits for which most parameters are the same, and only a
small subset vary. In order to handle these design constraints in the most
convenient for the user fashion, we have introduced command line options for
all parameters (all of which have ``reasonable'' default values) and an input
file for batch jobs, which lists only the parameters that change between
evolutions.

Since all options have default values, the simplest valid command line for
running \POET~is \verb#poet#. This will produce an output file called
poet.evol containing a pre--defined set of columns containing the evolution
for a one Jupiter mass planet around a solar mass star with default values
for all parameters needed for the evolution.

As noted in the main text, users can change the frequency dependence
of the tidal quality factor, as well as define custom stopping conditions.
The former requires the users to change a file named: \verb#StellarQ.cpp#
which contains the definitions of $Q_*(\omega_{orb}-\omega_{surf})$ and its
derivative. At present, $Q_*$ is not allowed to depend on anything except
the angular velocity of the tides as seen by the star. Defining custom
stopping conditions requires editing: \verb#ExternalStoppingConditions.h# and
\verb#ExternalStoppingConditions.cpp#. In either case, the user need not know
anything about the implementation details of \POET, but simply define the
relevant functions. Both types of modifications require re-compiling.

\section{Handling Stopping Conditions} \label{app: stopping conditions}
In \POET, stopping conditions are simply functions of the system age and
state that have a value of zero when the evolution should be stopped (either
because there is some discontinuity requiring a change in the differential
equations being solved, or because something that a user is interested in has
happened). They are assumed to be continuously differentiable at least up to
second order. Below we provide the full details of how \POET~handles stopping
conditions internally.

As the orbital evolution is being calculated, the values of the active
stopping conditions are stored, and when either a zero crossing or an
extremum for which the absolute value of any condition has a minimum is
detected, the evolution is forced to land on the zero--crossing or the
extremum to within some specified accuracy. 

The algorithm used is as follows:
\begin{enumerate}
	\item For a zero--crossing, if the absolute value of the stopping
		condition is smaller than some tolerance, the evolution is stopped
		and the appropriate changes are made to the equations and variables
		before it is continued. For an extremum, if the stopping condition
		value is within some tolerance of the estimated extremal value, and
		still no zero--crossing has been detected, the evolution simply
		continues. \label{stop step: check}
	\item If the last step did not take us close enough to the zero--crossing
		or the extremum, the evolution is reset to the last point before the
		event, the time of the extremum or zero--crossing is estimated and
		another step is taken to the estimated time. \label{stop step: try
		again}
	\item If the new point is before the zero--crossing or extremum, it is
		added to the stored evolution and the evolution is allowed to proceed
		(knowing it will very shortly be stopped again).
	\item If the new point is after the zero--crossing or extremum, we go
		back to step 1.
\end{enumerate}

Stopping conditions come in two flavors: those for which the first order
derivative is available and those for which it is not. In either case, a zero
crossing is detected by the change in sign of the stopping condition.

For stopping conditions with derivative information, extrema are detected by
a sign change in the first order derivative in two consecutive evolution
steps. Extrema are only investigated if the sign of the earlier derivative is
opposite the sign of the earlier stopping condition value. In this case, the
locations of zero--crossings and extrema, as well as values at the extremum
(needed for steps \ref{stop step: check} and \ref{stop step: try again}
above) are estimated from the unique third order polynomial that passes
through the two points surrounding the event and has the calculated
derivatives at those points.

For stopping conditions without derivative information, extrema are detected
and investigated by finding a time step for which the stopping condition
value is smaller in absolute value than for either of its neighbors. Without
derivative information, zero--crossing, extremum age and stopping condition
value are again estimated from a third order polynomial. The coefficients of
that polynomial in this case are calculated from four points in the evolution
that satisfy the following conditions:
\begin{itemize}
	\item they are either points from before the event, or are steps of
		various size started from the last point before.
	\item the zero--crossing or extremum being investigated occurred
		somewhere between the first and last point in the sequence.
	\item the time difference between the first and last points is the
		shortest possible given the above two constraints.
\end{itemize}

\begin{figure*}[t!]
	\begin{center}
		\includegraphics[width=\textwidth]{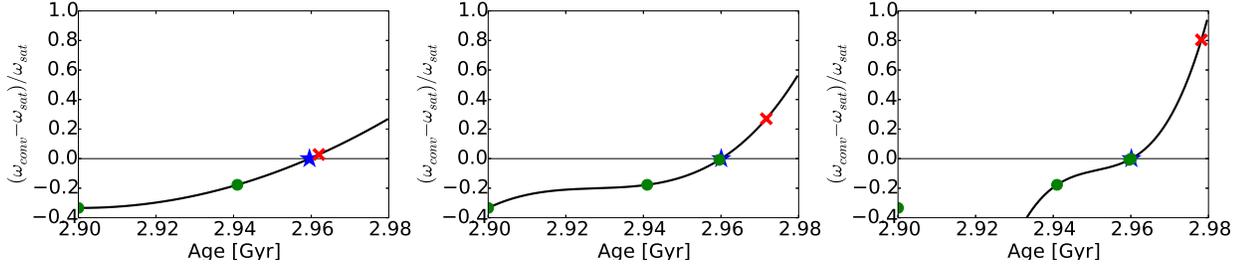}
		\caption{\footnotesize The wind saturation stopping condition at
		work (see text for details). Filled green circles indicate steps that
		are included in the evolution, red crosses indicate steps that are
		discarded, the blue star indicates the best estimate for the
		exact moment when the stopping condition is zero and the black curve
		shows the polynomial derived from some of the accepted and discarded
		steps to estimate the zero of the stopping condition.}
		\label{fig: stopping condition}
	\end{center}
\end{figure*}

Figure \ref{fig: stopping condition} demonstrates the workings of the wind
saturation stopping condition for the HAT-P-20 evolution shown in fig.
\ref{fig: HAT-P-20} to detect when the loss of angular momentum to the
stellar wind must be switched from its unsaturated to its saturated form (see
sec. \ref{sec: wind saturation modes}) as the star is spun--up in the final
stages of HAT-P-20b's inspiral. The stopping condition is defined as
$(\omega_{conv}-\omega_{sat})/\omega_{sat}$, where $\omega_{conv}$ is the
angular frequency at which the stellar convective zone spins, and
$\omega_{sat}$ is the frequency at which the magnetic wind saturates.

The panels in fig. \ref{fig: stopping condition} going from left to right
show consecutive steps taken by \POET~when calculating the evolution. In the
first panel, three steps after the last evolution mode change, the wind
saturation condition changes sign, from negative (the two green circles) to
positive (the red cross). This triggers the stopping condition mechanism.
Because only three points are available at this time, a second order
polynomial is derived that passes through the three points and is used to
estimate the time when the stopping condition is exactly zero (blue star).

In the middle panel, \POET~has taken a step of the appropriate size to land
on the estimated zero and found that the stopping condition is still negative
there (the right most green circle at an age of 2.96 Gyr). Because at this
point, the stopping condition has not changed sign yet, it is accepted and
the previous point which stepped over the sign change is discarded. Then,
another step is taken, resulting in a positive value of the wind saturation
condition (the red cross at an age of 2.97 Gyr). Now, four values for the
stopping condition are available, so a full third order polynomial is derived
passing through all of them, and a new estimate for the location of the zero
is derived (the blue star).

In the right panel, a step has been taken to the latest estimate for the
location of the zero, resulting in a negative stopping condition value, so it
is accepted and another step is taken yielding a  positive value. Now \POET~
has 5 points at its disposal to estimate the zero--crossing from (there are
actually two green circles very close to each other at an age of 2.96 Gyr),
but since only four can be used to derive the next estimate for the zero
crossing (no more than third order polynomials are used), the earliest point
(at an age of 2.90 Gyr) is not used. The new polynomial produces yet another
estimate for the zero--crossing, and this time when \POET~steps there, the
value of the stopping condition is zero to within the specified tolerance, so
the point is accepted, the evolution is interrupted and continues with the
saturated wind expression from then on.

\section{Stellar Evolution Interpolation}
\label{app: stellar evolution}

After experimenting with various algorithms for interpolating over a grid of
stellar evolution tracks the following prescription seemed to work best for
interpolating some quantity ($q$) to estimate the value it would take for a
star of mass $M_*$ at and age of $t$ ($q(M_*, t)$):
\begin{enumerate}
	\item Starting from a set of tracks calculated using the YREC model, for
		each track of mass $M_i$ a smoothing spline interpolation is derived
		(or is read from a previously serialized file) giving $q[M_i,
		\ln(t)]$, where $t$ is the stellar age. This happens at the beginning
		of the execution of a \poet~job.  \label{interp step: track splines}
	\item At each evaluation, for each stellar track corresponding to mass
		$M_i$ the pre--derived smoothing spline is evaluated to calculate
		$q_i\equiv q[M_i, \ln(t(M_*/M_i)^p)]$. By default $p=2.5$, but the
		value can be changed from the command line or from an input file (see
		appendix \ref{app: installation}).  \label{interp step: pick points}
	\item Derive a non-smoothing (passes exactly through the points) cubic
		spline of $q_i$ versus $M_i$ and evaluate it at $M_*$.  \label{interp
		step: mass interp}
\end{enumerate}

We scale the ages at which the evolutionary sequences in our YREC grids are
evaluated by the stellar mass ($M_i$) of each sequence because the key stages
in stellar evolution take place at different ages for stars of different
masses. We found that the particular scaling we use ($t(M_*/M_i)^{2.5}$) aligns
these stages for stars of different masses in a way that optimizes
interpolation.

\begin{figure*}[t!]
	\begin{center}
		\includegraphics[width=\textwidth]{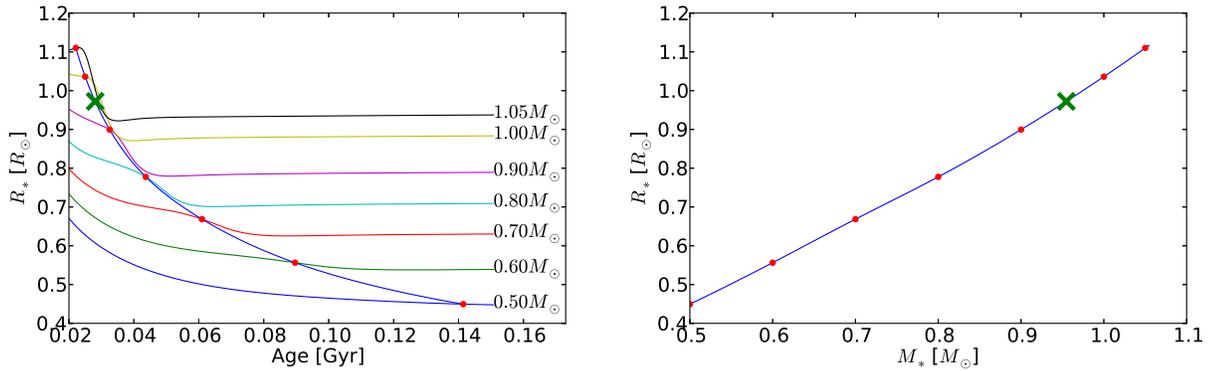}
		\caption{\footnotesize An illustration of how interpolating the stellar
		radius from the YREC tracks works. Left: the lines labeled with stellar
		masses are the smoothing cubic spline interpolations of the value of
		the radius ($R_*$) for each YREC track (step \ref{interp step: track
		splines}); the red circles show the set of points selected in step
		\ref{interp step: pick points}; the curve passing through all the red
		points is the spline derived by the mass interpolation, where we have
		assigned an age to each point scaled the same way as the track points
		were in step \ref{interp step: pick points}. Right: the red points
		are the same red points as in the left plot but plotted against mass
		instead of age; the curve passing through the points is the cubic
		spline derived in step \ref{interp step: mass interp}. In both plots,
		the green cross shows the final result of the interpolation.}
		\label{fig: stellar evolution interpolation}
	\end{center}
\end{figure*}

An illustration of how the stellar evolution interpolation works when
deriving the value of the stellar radius for a $0.95M_\odot$ star at an age
of 28.1 Myr is given in Figure \ref{fig: stellar evolution interpolation}.
The interpolation age was deliberately chosen to highlight the benefit of
scaling the ages at which tracks are evaluated. In the left panel of Fig.
\ref{fig: stellar evolution interpolation} we see that near that age the
radius of the star exhibits a sharp drop. By scaling the ages
at which each track is evaluated, we end up with always using the value of
the track right before the feature, leading to the very smooth curve shown in
the right panel of the figure. If instead we evaluated all tracks at the
desired age of 28.1 Myr, there would be a jump in the radius as a function of
mass between 0.9 and 1 $M_\odot$, which could lead to bad interpolation
results.

\end{document}